# RANDOM WALKS ON A FLUCTUATING LATTICE: A RENORMALIZATION GROUP APPROACH APPLIED IN ONE DIMENSION

C. D. Levermore
*Department of Mathematics*
*University of Arizona*
*Tucson, AZ 85721, USA*
lvrmr@math.arizona.edu

W. Nadler
*Institut für Theoretische Chemie*
*Universität Tübingen*
*Auf der Morgenstelle 8*
*D-72076 Tübingen, Germany*
ctina01@mailserv.zdv.uni-tuebingen.de

D. L. Stein
*Department of Physics*
*University of Arizona*
*Tucson, AZ 85721, USA*
dls@ccit.arizona.edu

### Abstract

We study the problem of a random walk on a lattice in which bonds connecting nearest neighbor sites open and close randomly in time, a situation often encountered in fluctuating media. We present a simple renormalization group technique to solve for the effective diffusive behavior at long times. For one-dimensional lattices we obtain better quantitative agreement with simulation data than earlier effective medium results. Our technique works in principle in any dimension, although the amount of computation required rises with dimensionality of the lattice.

## 1. Introduction

Many naturally occurring diffusive processes do not take place within a static medium. Influences of medium fluctuations on transport properties have been encountered particularly in polymeric host media (see [6] and references therein). To illustrate such a situation in the particular field of molecular biophysics, let us consider the migration of a ligand to a protein active site [13], a problem with which we have been concerned recently [9]. Several proteins (e.g. myoglobin) have their active site not on the surface, but inside the protein matrix. The ligand, in the case of myoglobin a small gas molecule like oxygen or carbon monoxide, has to reach its binding site, the heme pocket, by somehow crossing the protein matrix. An analysis of the conformational structure of myoglobin [4] suggests that no paths from the outside of the protein to the heme pocket are present when a protein is frozen into a static, average conformation. However, the protein conformational structure is not static — conformational fluctuations are present at physiological temperatures and, in fact, are crucial for the proper functioning of the protein [2,4]. There is a good deal of evidence that ligand diffusion cannot take place without local volume fluctuations within the protein. If myoglobin within a glycerol-water solvent, for example, is cooled to temperatures well below the glass transition of the solvent-protein system ($\approx 200°K$), then no ligand diffusion within the protein is observed [1]; presumably, a glass transition freezes out conformational fluctuations of the protein, at least on the timescales of the experiment, thereby prohibiting diffusion. Furthermore, the ligand almost certainly affects the neighborhood through which it is moving. Therefore, a correct treatment of ligand diffusion in myoglobin must take into account that local pathways for the ligand will appear and disappear randomly.

We are therefore led to consider the problem of macroscopic transport in a medium in which channels are randomly appearing and disappearing over a timescale $\tau$. A particular instance of such a situation is the case of a random walk that jumps from node to node on a lattice where the nearest-neighbor site connectivity fluctuates randomly in time. Simpler still, consider a lattice where each nearest-neighbor link opens and closes randomly, or, equivalently, where the nearest-neighbor hopping matrix element fluctuates between zero and some nonzero value. This last case, although special, is of great interest and wide applicability. A simplified version of this process has been used to study ionic conduction in polymeric solid electrolytes and protonic diffusion in hydrogen-bonded networks, among other things [6]. It is also clearly relevant to ligand diffusion in biomolecules such as hemoglobin [2], although a quantitative theoretical study of this situation has not been made.

For concreteness, first consider the static percolation problem ($\tau = \infty$) on a lattice in which a bond is present with probability $p$. Macroscopic transport (i.e. the ability of a particle to diffuse from one boundary of the system to another) requires that $p$ be greater than $p_c$, the percolation threshold of the lattice; $p_c = 1$ in one dimension, and depends on the lattice structure in higher dimensions. When $\tau < \infty$, macroscopic transport will be possible for any $p$. An "effective" diffusion constant $D_{eff}$ can be defined by $D_{eff} = \lim_{t \to \infty} \langle r^2(t) \rangle / t$, provided the limit exists, where $r(t)$ is the distance from the origin at time $t$ of a particle that starts at the origin at $t = 0$; $D_{eff}$ will depend on $p$ and $\tau$.

This problem has been analyzed in previous treatments [6,7,8,12,14], which used effective-medium theories for calculating $D_{eff}$. Qualitatively, their results agree with what one expects intuitively: For rapid bond fluctuations ($\tau$ small), $D_{eff}$ effectively tracks $p$, while for larger $\tau$ the dynamical problem approaches the static percolation problem, with an increasingly abrupt transition from low to high transport coefficient at the appropriate percolation threshold. However, although it is well known that effective medium theories can fail quantitatively [5], apparently no simulations have been performed up to now to check these results. If effective medium approaches fail to give satisfactory *quantitative* agreement with simulations, a new procedure which is capable of doing this is highly desirable, if one ultimately wishes to compare laboratory experiments (say, pressure and temperature studies of ligand diffusion in myoglobin) with theory.

The purpose of this paper is to propose a new treatment of the problem which does precisely that. It employs simple renormalization group ideas and is extremely easy to implement on a one-dimensional lattice. Furthermore, we compare both this implementation and earlier effective medium results with new, detailed numerical simulations.

## 2. The Model

To arrive at a better understanding of the processes discussed in the introduction, we will study a simplified problem in discrete space and time. Consider site diffusion on the lattice $\mathbb{Z}^d$. At each discrete timestep (which occurs at unit intervals) the particle must *attempt* to jump to a nearest neighbor site. There is *a priori* an equal probability of $1/2d$ for the particle to attempt to jump to any one of its $2d$ neighboring sites. This decision is made independently of the state of the bonds. Once the particle attempts to jump in a given direction, the jump will be successful if the relevant bond is present ("on"). If the bond is absent ("off"), the particle remains at its starting site until the next timestep, when again it attempts to jump in some direction. Let $b_{ij}(n)$ denote the state of the bond from site $j$ to site $i$ at timestep $n$ by $b_{ij}(n) = 0$ for off and $b_{ij}(n) = 1$ for on. Given a particular bond history

$$B = \{\{b_{ij}(0)\}, \cdots, \{b_{ij}(n)\}, \cdots\}, \qquad (2.1)$$

the particle dynamics can easily be cast into the following equation for the time evolution of the probability $P(i, n \mid B)$ that a random walker is at site $i$ at timestep $n$:

$$P(i, n+1 \mid B) = \frac{1}{2d} \left[ \sum_{\substack{j \in \mathbb{Z}^d \\ |j-i|=1}} b_{ij}(n) P(j, n \mid B) + P(i, n \mid B) \sum_{\substack{j \in \mathbb{Z}^d \\ |j-i|=1}} (1 - b_{ji}(n)) \right]. \qquad (2.2)$$

Note that the sums are over nearest neighbors. Our case is symmetric in the sense that $b_{ij}(n) = b_{ji}(n)$ holds; however, equation (2.2) can be readily used for the case of directed bonds, too. This jump process is Markov when the bond configuration is static because all attempts to jump are independent of earlier events. However, the jump process is not Markov when the bonds fluctuate because the bond state affects the jumps.

We now impose an independent bond dynamics on the bond variables $b_{ij}(n)$. Each bond independently fluctuates on and off at random times as a two-state Markov process that is identical for all bonds. The state of each bond at the discrete jumping times is then governed by the matrix of one-step transition probabilities $p_{b'b}$, the probability that $b_{ij}(n+1) = b'$ at timestep $n+1$ given that $b_{ij}(n) = b$ at timestep $n$. This transition matrix has the general form

$$\begin{pmatrix} p_{00} & p_{01} \\ p_{10} & p_{11} \end{pmatrix} = \begin{pmatrix} (1-p) + p\nu & (1-p)(1-\nu) \\ p(1-\nu) & p + (1-p)\nu \end{pmatrix}, \qquad (2.3)$$

where $0 \leq p \leq 1$ is the probability that any given bond is on and $0 \leq \nu \leq 1$ is the correlation factor. Iterating the transition matrix $n$ times then gives the matrix of $n$-step transition probabilities $p_{b'b}(n)$ to be

$$\begin{pmatrix} p_{00}(n) & p_{01}(n) \\ p_{10}(n) & p_{11}(n) \end{pmatrix} \equiv \begin{pmatrix} p_{00} & p_{01} \\ p_{10} & p_{11} \end{pmatrix}^n = \begin{pmatrix} (1-p) + p\nu^n & (1-p)(1-\nu^n) \\ p(1-\nu^n) & p + (1-p)\nu^n \end{pmatrix}. \qquad (2.4)$$

This shows that when $\nu < 1$ temporal correlations decay as $\nu^n$. The Markov property allows us to consider the bond dynamics to be a continuous time Poisson process without loss of generality. In this case the correlation time $0 \leq \tau \leq \infty$ (in lattice temporal units) is determined by $\nu = \exp(-1/\tau)$ and the mean lengths of time over which a bond remains either on or off, denoted $\tau_{on}$ and $\tau_{off}$, are given by $\tau_{on} = \tau/(1-p)$ and $\tau_{off} = \tau/p$ respectively. When $\nu = 1$ ($\tau = \infty$) the transition matrix (2.3) becomes the identity matrix and the bonds become static. When $\nu = 0$ ($\tau = 0$) the bonds are completely de-correlated from one jumping time to the next. Because the bond process is Markov, the joint bond/jumping process is also Markov.

Previous work [7,10] has indicated that at long times the behavior of this and related models is diffusive for $0 < p$ and $\nu < 1$ (also see Figure 4 below), namely, it was found that there exists a positive constant $D_{eff}$ such that the relation

$$\langle r^2(n) \rangle \approx D_{eff}\, n \qquad (2.5)$$

holds for large values of $n$. These results strongly suggest, and we will henceforth assume, that this effective diffusion coefficient exists and is defined by the limit

$$D_{eff} \equiv \lim_{n \to \infty} \frac{\langle r^2(n) \rangle}{n} \equiv \lim_{n \to \infty} \frac{1}{n} \sum_{i \in \mathbb{Z}^d} |i|^2 P(i, n), \qquad (2.6)$$

where $P(i, n)$ is the probability of finding a particle at the site $i$ at timestep $n$ given that it started at the origin at time 0 and that $|i|$ is the distance of site $i$ from the origin. Thus, $P(i, n)$ is the expected value of $P(i, n \,|\, B)$ over all bond histories $B$, where $P(i, n \,|\, B)$ is computed for a given $B$ as the solution of (2.2) that satifies the initial condition $P(i, 0 \,|\, B) = \delta(i)$. Here $\delta(\cdot)$ denotes the Kronecker delta function centered at the origin. It is clear that the $P(i, n)$, and hence $D_{eff}$, depend on $(p, \nu)$, the parameters associated with the ensemble of bond histories $B$. This dependence will sometimes be indicated by writing $P(i, n \,|\, p, \nu)$ and $D_{eff}(p, \nu)$. While it may be that $P(i, n)$ is non-Gaussian [10], we will not address that question here, but rather study only the determination of the effective diffusion coefficient $D_{eff} = D_{eff}(p, \nu)$.

The relation of the $D_{eff}$ defined above in (2.6) to a continuum diffusion description can be understood by considering the particle dynamics on macroscopic scales in which the unit lattice spacing and unit timestep increment have sizes designated $\delta x$ and $\delta t$ respectively. Formally taking a macroscopic limit in which $\delta x$ and $\delta t$ vanish while $(\delta x)^2/\delta t$ is held fixed, the macroscopic density of particles $\rho = \rho(t,x)$ will satisfy the diffusion equation

$$\partial_t \rho = \frac{(\delta x)^2}{2d\delta t} D_{eff} \Delta_x \rho \,. \tag{2.7}$$

The continuum diffusion coefficient is therefore proportional to both $D_{eff}$ and the dimensional ratio $(\delta x)^2/\delta t$, which is invariant under the spatio-temporal scale symmetry ($x \to ax$, $t \to a^2 t$) of the equation.

The value of $D_{eff}(p,\nu)$ may be easily determined in the following four limiting cases. First, when $p = 1$ the bonds are always on and the particle dynamics reduces to free diffusion with $D_{eff} = 1$. Because bonds being off can only reduce the transport of particles, it follows that one must generally have $D_{eff} \leq 1$, with inequality indicating deviations from free diffusion. Second, when $p = 0$ the bonds are always off and clearly $D_{eff} = 0$. Third, when $\nu = 0$ the bonds fluctuate infinitely fast and the expected state of the bonds is independent from timestep to timestep. In that case it is easy to show that $D_{eff} = p$. Finally, when $\nu = 1$ the problem becomes that of diffusion on a static, randomly bond-diluted lattice. In this case, the behavior will strongly depend on the dimension $d$ of the lattice. For $d = 1$ it is easily seen that

$$D_{eff}(p,1) = \begin{cases} 1 & \text{if } p = 1 \,, \\ 0 & \text{otherwise} \,. \end{cases} \tag{2.8}$$

For $d > 1$ the behavior will be diffusive only if $p > p_c$, where $p_c < 1$ is the bond percolation threshold for the lattice in question, which depends on $d$. It is well-known that in this situation diffusion is anomalous for intermediate times and the limit (2.6) is approached only for very long times; exactly at $p_c$ diffusion is anomalous for all times [3,11].

In addition to the above limiting cases, it may be readily argued that $(p,\nu) \mapsto D_{eff}(p,\nu)$ must be an increasing function of $p$ and a decreasing function of $\nu$. However, we would like to know quantitatively the full range of behavior of $D_{eff}$ as a function of $p$ and $\nu$. We have developed a renormalization group scheme that provides a new, non-effective medium method for computing this, which will be the subject of the following section.

## 3. The Renormalization Group Procedure

Renormalization group (RG) transformations are best understood as transformations in the parameter space of the models in question that are connected with a rescaling of space and/or time, and that leave macroscopic properties of the system invariant. For the fluctuating bond lattices introduced in the last section the exact macroscopic dynamics would be recovered if we could find an RG procedure that coarsens the spatial and temporal units of the lattice so that the ratio $(\delta x)^2/\delta t$ remains fixed while the parameters $p$ and $\nu$ are transformed so that the macroscopic bond fluctuation correlation time $\tau \delta t$ and the effective diffusion coefficient $D_{eff}$ are invariant. Of course, we cannot fix $D_{eff}$ exactly because that would require prior knowledge of $D_{eff}$. Rather, what we will do is to match one of the approximates to $D_{eff}$ from the limiting relation (2.6) for the fine lattice with the appropriate approximate for the coarse lattice. Upon iterating the resulting RG transformation to a fixed point we will obtain an approximation to $D_{eff}$ which will prove remarkably accurate in one dimension.

Let $D_n(p,\nu)$ denote the $n^{th}$ approximate to $D_{eff}(p,\nu)$ in (2.6), which has the form

$$D_n(p,\nu) \equiv \frac{\langle r^2(n) \rangle}{n} = \frac{1}{n} \sum_{i \in \mathbb{Z}^d} |i|^2 P(i, n \,|\, p, \nu) \,. \tag{3.1}$$

The $P(i, n \,|\, p, \nu)$, and hence $D_n(p,\nu)$, can be determined explicitly for any value of $n$. This task is relatively easy when $n$ is small. For example, when $n = 1$ a particle can have only either moved to a nearest neighbor or remained at the origin. Because the probability of attempting to move in a given direction is $1/2d$ while the probability of the move being successful is $p$, one has

$$P(i, 1 \,|\, p, \nu) = \begin{cases} (1-p) & \text{if } |i| = 0 \,, \\ \dfrac{1}{2d} p & \text{if } |i| = 1 \,, \\ 0 & \text{otherwise} \,, \end{cases} \tag{3.2}$$

which by (3.1) gives

$$D_1(p,\nu) = \sum_{i \in \mathbb{Z}^d} |i|^2 P(i, 1 \,|\, p, \nu) = \sum_{\substack{i \in \mathbb{Z}^d \\ |i|=1}} \frac{1}{2d} p = p \,. \tag{3.3}$$

Similarly, when $n = 2$ a particle can have only either moved to a next-nearest neighbor, moved to a nearest neighbor or remained at the origin. By examining the likelihood of each possible path a particle could take to end up at $i$ after two timesteps, one can show

$$P(i, 2 \,|\, p, \nu) = \begin{cases} \dfrac{1}{2d} p^2 + (1-p)^2 + \dfrac{1}{d} p(1-p)\nu & \text{if } |i| = 0 \,, \\ \dfrac{1}{2d^2} p(1-p)(2d-\nu) & \text{if } |i| = 1 \,, \\ \dfrac{1}{2d^2} p^2 & \text{if } |i| = \sqrt{2} \,, \\ \dfrac{1}{4d^2} p^2 & \text{if } |i| = 2 \,, \\ 0 & \text{otherwise} \,, \end{cases} \tag{3.4}$$

which by (3.1) gives

$$D_2(p,\nu) = \frac{1}{2} \sum_{i \in \mathbb{Z}^d} |i|^2 P(i, 2 \,|\, p, \nu) = p - \frac{1}{2d} p(1-p)\nu \,. \tag{3.5}$$

It is already clear that the complexity of the calculation of the $P(i, n \,|\, p, \nu)$ increases rapidly with $n$. Later we will describe a method for carrying out these calculations for general $n$.

Now we consider a few general properties of $D_n$. First, each $D_n(p,\nu)$ has some of the same limiting behaviors enjoyed by $D_{eff}$, namely that $D_n$ is an increasing function of $p$ and a decreasing function of $\nu$ such that

$$D_n(p,0) = p\,, \qquad D_n(0,\nu) = 0\,, \qquad D_n(1,\nu) = 1\,, \qquad D_n(p,\nu) \leq p\,, \tag{3.6}$$

with equality only for $\nu = 0$ or $p = 0, 1$ when $n > 1$. Second, as $\nu$ is decreased the variation of $D_n(p,\nu)$ with $n$ becomes smaller and the convergence of (2.6) becomes more rapid.

We now assume that an evaluation of $D_n(p,\nu)$ is given for some fixed $n > 1$ and introduce an $n$-step RG transformation of the model parameters $(p,\nu)$ to new values $(p',\nu')$ that corresponds to a coarsening of the spatial and temporal units of the lattice by factors of $\sqrt{n}$ and $n$ respectively. The $n$-step RG transformation is chosen to match the bond $n$-step correlation factor and $n^{th}$ approximation to $D_{eff}$ for the original system with the one-step correlation factor and first approximation to $D_{eff}$ for the new system. Specifically, this means that by (2.4)

$$\nu' = \nu^n\,, \tag{3.7a}$$

thereby fixing the macroscopic bond fluctuation correlation time, while by (3.3)

$$p' = D_1(p',\nu') = D_n(p,\nu)\,. \tag{3.7b}$$

It is seen immediately that these equations already represent the RG transformation equations for $p'$ and $\nu'$ which introduce a flow in the parameter space of $(p,\nu)$, see Figure 1. From the properties (3.6) of $D_n(p,\nu)$, one can easily see that $(p,\nu) = (0,1)$ and $(p,\nu) = (1,1)$ are unstable fixed points, and $\nu = 0$ is a line of stable fixed points under the above RG transformation.

Our strategy for the approximation of $D_{eff}(p,\nu)$ is now straightforward: given $n > 1$, we start with the bare model parameters $(p,\nu)$ and iterate the $n$-step RG transformation (3.7). Whenever $(p,\nu)$ is not $(0,1)$ or $(1,1)$, this process will approach a fixed point of the form $(p^*, 0)$, in which case we assign

$$D_{eff}^{RG}(p,\nu) \equiv D_{eff}(p^*,0) = p^*\,. \tag{3.8}$$

This value for the effective diffusion coefficient is associated with all models that are connected by the same RG trajectory. However, it will be far more accurate for those values of $\nu$ that are near zero, where the convergence of (2.6) is rapid, rather than those near 1. This strategy for computing $D_{eff}^{RG}(p,\nu)$ is carried out in the next section for a one-dimensional lattice and for different values of $n$, and the results are analyzed and compared with simulations.

## 4. The RG for a One-Dimensional Lattice

We now specialize the RG procedure to a one-dimensional lattice, deriving the two-, three-, and four-step RG transformations. In the next section we will demonstrate that the almost trivial two-step renormalization given below is sufficient to give very good agreement with numerical simulations, better in fact than effective medium theories which require considerably more work. We will improve upon this agreement by continuing to the three- and four-step renormalization.

It should be noted that the number of terms, and hence the labor required to compute the RG transformation, increases exponentially with the number of steps used.

The $n$-step renormalization requires knowledge of $D_n(p,\nu)$, which is computed from the $P(i,n\,|\,p,\nu)$ by formula (3.1). For the one-dimensional problem the $i$ runs over $\mathbb{Z}$. Fixing $(p,\nu)$ and noting the general symmetry $P(-i,n) = P(i,n)$, the probabilities $P(i,n)$ need only be computed for $i = 1, \cdots, n$. Of course, the probability $P(0,n)$ can be determined from the others by the relation

$$P(0,n) = 1 - 2\sum_{i=1}^{n} P(i,n), \tag{4.1}$$

but this is not necessary because of its vanishing contribution to (3.1). Below we will use the two-step case to illustrate how the $P(i,n)$ can be calculated using a diagrammatic technique to classify the various possible paths.

Although it is not necessary for this calculation, the diagrammatic method introduced here for counting particle paths is helpful when more extensive computations are required. The diagrams we use are a shorthand for collecting related terms. Consider the diagram shown in Figure 2. Each level downward corresponds to a new increment of time. At the top level there is only one site shown, indicating that at time $m = 0$ the particle is at the origin; one level down ($m = 1$) three sites are shown, indicating that the particle can be at sites $-1$, $0$, or $1$; and two levels down ($m = 2$) five sites are shown ($i = -2$ through $i = 2$, inclusive). Our diagrams then trace possible particle paths, with the understanding that at time zero the particle starts at the origin.

All two-step paths are indicated in diagrams ($a - f$) of Figure 3. In addition to $P(2,2)$ and $P(1,2)$, we will compute the probability $P(0,2)$ for illustrative purposes. When the final position of the particle is at site $i$ the contribution of each diagram to the calculation of $P(i,2)$ is as follows:

$i = 2$: (Diagram $a$) The particle goes right twice (which has a probabilistic weight of $\frac{1}{4}p^2$).

$i = 1$: (Diagram $b$) There are two possibilities for this diagram: the particle first fails to go right and then goes right (which has weight $\frac{1}{4}(1-p)p_{10}$); the particle first fails to go left and then goes right (weight $\frac{1}{4}(1-p)p$). The total weight for this diagram is $\frac{1}{4}(1-p)(p_{10}+p)$.

(Diagram $c$) There are two possibilities for this diagram: the particle first goes right and then fails to go left (weight $\frac{1}{4}p\,p_{01}$); the particle first goes right and then fails to go right again (weight $\frac{1}{4}p(1-p)$). The total weight for this diagram is $\frac{1}{4}p(1-p+p_{01})$.

$i = 0$: (Diagram $d$) The particle first goes right and then goes left (weight $\frac{1}{4}p\,p_{11}$).

(Diagram $e$) The particle first goes left and then goes right (weight $\frac{1}{4}p\,p_{11}$).

(Diagram $f$) There are four possibilities for this diagram: the particle fails to go right twice (weight $\frac{1}{4}(1-p)p_{00}$); the particle fails to go left twice (weight $\frac{1}{4}(1-p)p_{00}$); the particle first fails to go right and then fails to go left (weight $\frac{1}{4}(1-p)^2$); the particle first fails to go left and then fails to go right (weight $\frac{1}{4}(1-p)^2$). The total weight for this diagram is $\frac{1}{2}(1-p)(1-p+p_{00})$.

Summing the weights of the appropriate diagrams and eliminating the transition probabilities $p_{b'b}$ using definition (2.3) then gives

$$P(2,2\,|\,p,\nu) = \tfrac{1}{4}p^2\,,$$
$$P(1,2\,|\,p,\nu) = \tfrac{1}{2}p(1-p)(2-\nu)\,, \qquad (4.2)$$
$$P(0,2\,|\,p,\nu) = \tfrac{1}{2}p^2 + (1-p)^2 + p(1-p)\nu\,,$$

which agrees with (3.4) when $d = 1$. The verification of (4.1) by these probabilities provides a useful check of the calculation. Using the probabilities (4.2) to evaluate the second approximation to $D_{eff}$ by (3.1) then yields

$$D_2(p,\nu) = P(1,2\,|\,p,\nu) + 4P(2,2\,|\,p,\nu) = p - \tfrac{1}{2}p(1-p)\nu\,. \qquad (4.3)$$

The approximate effective diffusion constant $D_{eff}^{RG}$ is then found by iterating the two-step RG transformation (3.7). In particular, it is interesting to note that when $\nu = 1$ this procedure recovers (2.8) exactly.

A better value of $D_{eff}^{RG}$ should be obtained when the RG transformation is based on a higher order approximation to $D_{eff}$. Applying the above diagrammatic technique to compute all possible particle paths, for the three-step case we found

$$P(3,3\,|\,p,\nu) = \tfrac{1}{8}p^3\,,$$
$$P(2,3\,|\,p,\nu) = \tfrac{1}{4}p^2(1-p)(3-2\nu)\,, \qquad (4.4)$$
$$P(1,3\,|\,p,\nu) = \tfrac{3}{8}p\bigl(p^2 + 4(1-p)^2\bigr) - p(1-p)(1-2p)\nu - \tfrac{1}{4}p^2(1-p)\nu^2\,,$$

while for the four-step case we found

$$P(4,4\,|\,p,\nu) = \tfrac{1}{16}p^4\,,$$
$$P(3,4\,|\,p,\nu) = \tfrac{1}{8}p^3(1-p)(4-3\nu)\,,$$
$$P(2,4\,|\,p,\nu) = \tfrac{1}{4}p^2\bigl(6 - 12p + 7p^2\bigr) - \tfrac{3}{4}p^2\bigl(2 - 5p + 3p^2\bigr)\nu\,, \qquad (4.5)$$
$$P(1,4\,|\,p,\nu) = \tfrac{1}{2}p\bigl(4 - 12p + 15p^2 - 7p^3\bigr) - \tfrac{3}{8}p\bigl(4 - 20p + 31p^2 - 15p^3\bigr)\nu$$
$$\qquad\qquad - \tfrac{1}{2}p^2\bigl(3 - 7p + 4p^2\bigr)\nu^2 - \tfrac{1}{4}p^3(1-p)\nu^3 - \tfrac{3}{8}p^2(1-p)^2\nu^4\,.$$

The third and fourth approximations to $D_{eff}$ are then found by formula (3.1) to be

$$D_3(p,\nu) = p - \tfrac{2}{3}p(1-p)\nu - \tfrac{1}{6}p^2(1-p)\nu^2\,,$$
$$D_4(p,\nu) = p - \tfrac{3}{4}p(1-p)\nu - \tfrac{1}{4}p^2(1-p)\nu^2 - \tfrac{3}{8}p^3(1-p)\nu^3 - \tfrac{1}{16}p^2(1-p)^2\nu^4\,. \qquad (4.6)$$

Because the procedures to calculate these cases are essentially no different from those for the two-step case, we omit the details. The effective diffusion constants are again found by iterating the three-step and four-step RG transformations (3.7).

## 5. Comparisons with Monte Carlo Simulations

Extensive Monte Carlo (MC) simulations of random walks in a one-dimensional fluctuating bond system were performed. We placed $N$ non-interacting walkers randomly on a chain of $L$ sites connected by fluctuating bonds where periodic boundary conditions were employed. At each time step the state of each lattice bond was updated in accordance with (2.3), then each walker attempted a move to a neighboring site in a random direction. The move was accepted if the bond connecting the sites in question was on. This scheme corresponds to the analytic model described in Section 2. The simulation was stopped when a walker reached a distance of $L/2$ from its starting point, in order to avoid problems due to the toroidal topology introduced by the periodic boundary conditions. For our simulations we usually chose $N = L$ walkers, so that each lattice site was occupied with one walker on average. Averaging the square of the displacements of these random walkers from their starting positions over $M$ runs provides us with averages over walks as well as over bond fluctuations in computing $\langle r^2(n) \rangle$. Typically the lattice size $L$ employed was $10^3$ and the number of runs $M$ ranged from 10 to 100, depending on the quality of the statistics. The results were checked against invariance with respect to variations in the parameters $L, M$, and $N$.

A typical result for the mean square displacement $\langle r^2(n) \rangle$ is shown in Figure 4. The data can be described by

$$\langle r^2(n) \rangle = D_{eff} n + r_0^2 \left(1 - \phi(n)\right) \tag{5.1}$$

where $D_{eff}$ is the effective diffusion coefficient, $r_0^2$ is a residual mean square displacement, and $\phi(n)$ is a — usually non-exponential — relaxation function with $\phi(0) = 1$ and $\lim_{n \to \infty} \phi(n) \to 0$.

In Figure 5 the RG and MC-simulation results for $D_{eff}$ are compared; since the small $p$ regime cannot be resolved readily when $D_{eff}$ is plotted linearly, the data are replotted in Figure 6 with a logarithmic scale for $D_{eff}$. As can be seen, already the two-step RG transformation gives a reasonable agreement with the simulation results, which is improved upon increasing the step size. In order to give a more quantitative description of the agreement, we have analyzed the relative error $(D_{eff}^{RG} - D_{eff}^{sim})/D_{eff}^{sim}$. Figure 7 demonstrates the decrease of the relative error with step size for $\nu = .999$. However, as can be seen, the convergence to zero error is rather slow. Surprisingly, the relative error is largest (about 50%) for small values of $p$, whereas the RG gives correct results for $p = 0$. In addition, Figure 8 demonstrates that the relative error is rather independent of $\nu$, particularly for small $p$.

It is interesting to compare these results with the prediction of effective medium (EM) theories. These theories are usually expressed in terms of the correlation time $\tau$, which is related to $\nu$ by

$$\frac{1}{\tau} = \log\left(\frac{1}{\nu}\right). \tag{5.2}$$

We note that various EM approaches give somewhat different predictions; e.g. [14] predicts a scaling $D_{eff} \propto p^2/\tau$ for small $p$ and large $\tau$, whereas [7] predicts $D_{eff} \propto p/\tau$, in agreement with our MC-simulations and RG results. Therefore we choose to compare the results of [7] with our simulations. The one-dimensional self-consistency equation of [7] can be evaluated analytically

with the result

$$D_{eff}^{EM} = 1 + 2\tau(1-p)^2 - \sqrt{\left(1 + 2\tau(1-p)^2\right)^2 - p(2-p)}. \tag{5.3}$$

Note that this equation gives the correct results for $p = 0, 1$ and for $\tau = 0$ (alternatively, for $\nu = 0$ by (5.2)). Figure 9 shows the relative error of this prediction for $D_{eff}$ with respect to our simulations. Notice that EM theory systematically underestimates the effective diffusion coefficient, the error increasing with $\tau$, particularly in the intermediate to large $p$ regime, i.e., below the percolation threshold in one dimension. Everywhere, the relative error is larger even than that of the two-step RG results.

## 6. Discussion

We have presented a renormalization group approach to obtaining an approximate effective diffusion coefficient for random walks on a fluctuating lattice. This procedure was applied to a one-dimensional lattice, where it is relatively simple to implement, and was found to be in good quantitative agreement with Monte Carlo simulations. The results can be improved by taking into account a larger step size in the renormalization procedure. It might be hoped that, e.g., using a multi-step renormalization scheme, the procedure presented could be modified in order to account also for nondiffusive effects; see (5.1).

The application of this approach to higher dimensions — two is already interesting — is straightforward, but requires much more calculation than the one-dimensional problem considered above. In addition, there appears to arise a problem of principle: It is reasonable to expect that, as $\nu \to 1$, one should be able to recover the percolation limit. Specifically, as $\nu \to 1$ while holding $p$ fixed, we expect the limiting diffusion coefficient as a function of $p$ should vanish for $p < p_c$, and be positive for $p > p_c$. This behavior should be reflected in a renormalization group flow like it is sketched in Figure 10. In particular, there should arise a nontrivial fixed point at $\nu = 1$ and $p = p_c$. However, from the general properties (3.6) of the $D_n$ one can immediately conclude that such a fixed point cannot arise for any finite $n$ in our renormalization procedure. So, the best that can be hoped for is that the proper behavior is approached as the number of steps in the renormalization goes to infinity. Said another way — because of the dimensional dependence of $p_c$, one must take enough steps in the renormalization procedure to "see" the dimensionality of the lattice. The two-step procedure, employed so successfully in one dimension, does not give a very good approximation in two dimensions. However, the analogue of the four-step procedure, which is the minimum needed to see simple closed paths, begins to see the proper trend.

**Acknowledgements**

The work of C.D.L. was partially supported by the AFOSR under grant F49620-92-J-0054 at the University of Arizona. The work of D.L.S. was partially supported by the DOE under grant DE-FG03-93ER25155 at the University of Arizona. The work of D.L.S. and W.N. was also partially supported by a NATO Collaborative Research Grant. Some of this work was carried out while C.D.L. was visiting the Mathematical Sciences Research Institute (MSRI) in Berkeley, which is supported in part by the NSF under grant DMS-9022140.

**Figure Captions**

Fig. 1: Renormalization group flow for $d = 1$; crosses denote fixed points.

Fig. 2: A nodal tree for $n = 2$.

Fig. 3: Diagams $a - f$ showing all the possible paths ending at $i = 2, 1, 0$.

Fig. 4: Average mean square displacement $\langle r^2(n) \rangle$ vs $n$; parameters as indicated; the dotted line denotes the long-time behavior, see (5.1).

Fig. 5: Comparison of 2-step (alternating lines), 3-step (dashed lines), and 4-step (solid lines) RG results, and MC-simulations for $D_{eff}$ vs $p$ for $\nu = 0, 0.9$ (circles), $0.99$ (squares), $0.999$ (triangles).

Fig. 6: Same as Fig.5, with $D_{eff}$ on a logarithmic scale in order to resolve the small $p$ regime.

Fig. 7: Comparison of the relative error of the 2,3,and 4-step RG results vs $p$; $\nu = 0.999$.

Fig. 8: Relative error vs $p$ of the 4-step RG results for $\nu = 0.9, 0.99, 0.999$.

Fig. 9: Relative error vs $p$ of effective medium results for $\nu = 0.9, 0.99, 0.999$.

Fig. 10: Renormalization group flow for $d > 1$; crosses denote fixed points.

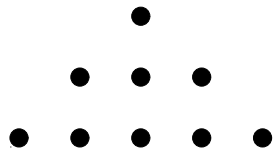

**Figure 2:** A nodal tree for $n = 2$.

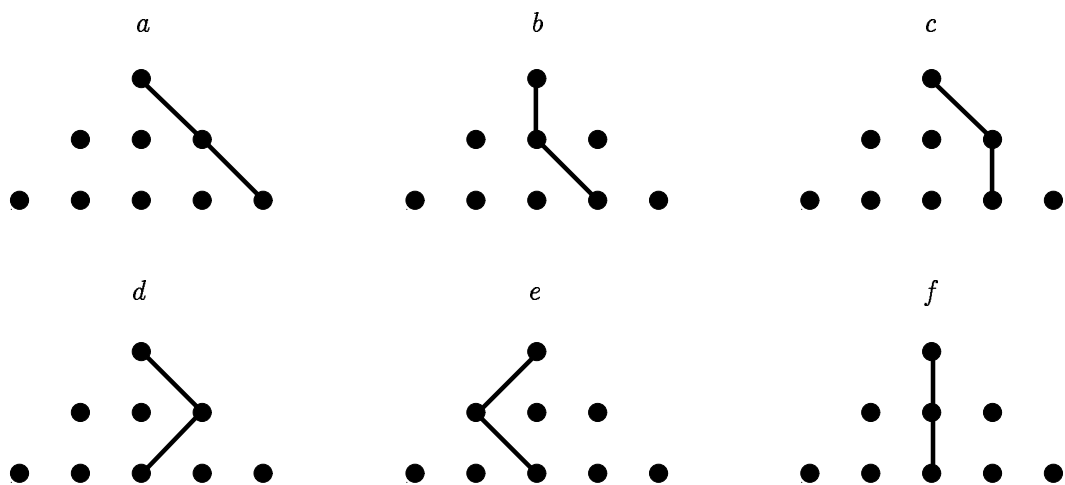

**Figure 3:** Diagrams $a - f$ showing all the possible paths ending at $i = 2, 1, 0$.